# Kinetic Model for a Threshold Filter in an Enzymatic System for Bioanalytical and Biocomputing Applications

Vladimir Privman,[a]\*  Sergii Domanskyi,[a]  Shay Mailloux,[b]
Yaovi Holade,[b,c]  Evgeny Katz[b]\*\*

[a]*Department of Physics, and*
[b]*Department of Chemistry and Biomolecular Science, Clarkson University, Potsdam, NY 13676*
[c]*Université de Poitiers, IC2MP, UMR-CNRS 7285, 4 rue Michel Brunet, B27 TSA 51106, 86073 Poitiers Cedex 9, France*

Corresponding authors' contact information:

\*privman@clarkson.edu, http://www.clarkson.edu/Privman, +1-315-268-3891

\*\*ekatz@clarkson.edu, http://www.clarkson.edu/~ekatz, +1-315-268-4421

## Abstract

A recently experimentally observed biochemical "threshold filtering" mechanism by processes catalyzed by the enzyme malate dehydrogenase is explained in terms of a model that incorporates an unusual mechanism of inhibition of this enzyme that has a reversible mechanism of action. Experimental data for a system in which the output signal is produced by biocatalytic processes of the enzyme glucose dehydrogenase are analyzed to verify the model's validity. We also establish that fast reversible conversion of the output product to another compound, without the additional inhibition, cannot on its own result in filtering.

# INTRODUCTION

Recent research in signal processing in molecular and biomolecular systems has focused on novel applications that involve cascades of chemical or biochemical reactions with several reagents as input/output signals.[1-14] Such multi-input reaction cascades are used in biosensing, biomolecular computing, and decision making devices and setups utilizing (bio)chemical processes with well-defined responses.[15-20] These systems require experimental design and theoretical understanding[21-23] of chemical and biochemical reactions that allow signal response modification and control. Such processes are then incorporated in more complicated biochemical "networks" of concatenated reactions.[24-29] Scheme 1(a) shows a typical response of a (bio)chemical process of input (at time $t = 0$) to output (at a later "gate" time $t = t_g$) transduction, with approximately linear behavior at low inputs followed by a saturation region at larger inputs. The former is due to a low supply of the input, whereas the latter usually originates because other reactants limit the process rate.

In many biosensing applications it is useful to modify such a generic response to make it as linear as possible,[30-44] by converting the response shape of Scheme 1(a) to that shown in Scheme 1(b), without too much loss of the over signal intensity. Recent work has involved[30,31] data analysis to develop a theoretical understanding of how two enzymatic processes with different nonlinear responses can be combined to yield an extended linear response regime. Multi-step and multi-input signal processing has been studied as potentially enabling biomolecular computing,[5-9,45] i.e., information processing involving cascades of (bio)chemical reactions rather than being solely based on electronics. Biomolecular computing is a sub-field of unconventional computing,[46-48] and it requires a "toolbox" of logic elements in order to build a binary network. Various binary logic gates have been demonstrated, including AND, OR, NAND, NOR, CNOT, XOR, INHIBIT, etc.[6-9] Such gates were also connected in small networks,[24-29] some if which carried out simple computational[49,50] and information storage[51,52] steps.

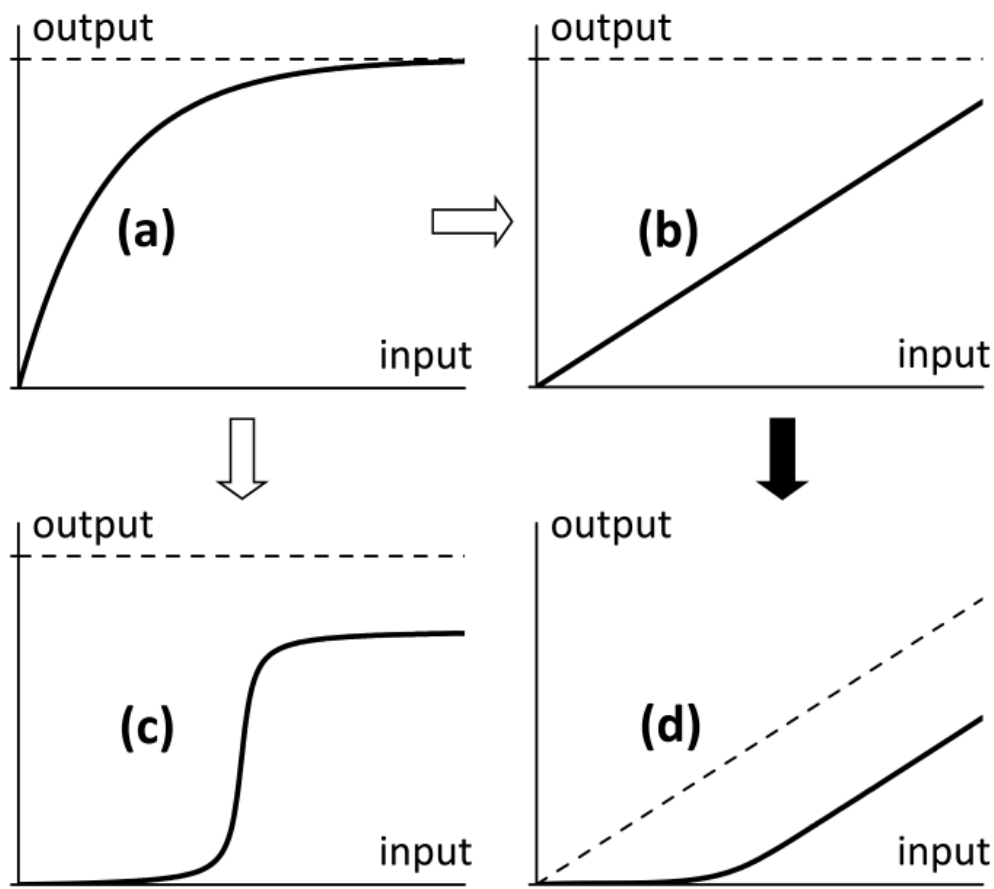

**Scheme 1.** (a) A typical "convex" response shape for a chemical or biochemical process used as a single-input to single-output "transduction" step in signal/information processing. (b) Linear response desirable in many biosensing applications. The conversion from the convex to linear response, (a) ⇒ (b), can result in the loss of some of the overall signal intensity. (c) "Binary" sigmoid-shape response of interest in biomolecular computing. The desirable response function should be symmetrical and steep at the middle inflection. The (a) ⇒ (c) conversion also typically results in the loss of some intensity. (d) In applications of interest, the conversion of a linear response to the threshold one, followed by a linear behavior, (b) → (d), is required. In this case the goal is to have a well-defined (practically zero slope) pre-threshold region for low inputs, and to preserve the sensitivity (the slope of the output curve) for large inputs.

Both for individual binary two-input single-output gate realizations and especially for their networking, it is useful to have added non-binary network steps that convert a convex response, Scheme 1(a), to sigmoid, such as illustrated in Scheme 1(c). The goal for the simplest single input transduction is to have the resulting sigmoid response as symmetric and steep as possible, preserving the saturation regime at large inputs, and without losing too much of the overall signal intensity. Most of the reported designs and realizations[31,53-65] of such "biochemical filtering" involved modifying the response at low inputs to a practically zero-slope, while largely preserving the saturation regime. These approaches, including generalizations to two-input processes, typically involve "intensity filtering" whereby the input or output is efficiently diverted (chemically consumed), but only up to a limited quantity. The added "filtering" process thus only affects (dampens) the low-input/low-output response. However, such systems usually do lose some overall signal intensity and have a less well-defined saturation regime at large inputs.

The work on binary logic has used biomolecular processes involving proteins/enzymes,[9,66] DNA/RNA,[5-8,67-69] and bacterial cells.[70,71] Enzymatic processes are of great interest because they promise short-term development of new biosensing[19,20,24,27,72-74] and bioactuating applications[75-77] with even moderate-complexity reaction cascades. Indeed, most biosensing and bioanalytical devices involve enzymatic reactions, which are also relatively easy to integrate with electronics.[78] The enzyme-based logic systems operating as binary (YES/NO) biosensors were also interfaced with electrochemical/electronic devices represented by electrodes[73] or field-effect transistors.[18,79,80]

Recently, experiments[25] on three-input majority and minority enzymatic gates for biocomputing applications have underscored the importance of another type of "biochemical filtering" as a part of the output biochemical post-processing to achieve the desired response. In this case the response of Scheme 1(b) is converted to that of Scheme 1(d), with the goal of controlling the threshold, i.e., the offset region of small inputs for which the output remains practically zero.

In this work we demonstrate that the added "filtering" mechanism in the referenced experiments[25] (and in the earlier work on filtering[81]) utilizing the enzyme malate dehydrogenase (malic dehydrogenase), MDH, is based on an unusual mechanism of enzymatic biocatalytic activity of this enzyme, noted in an early work on the mechanism of action of MDH.[82] This work[82] considered what is called[83] a reversible random-sequential bi bi mechanism of action for MDH, and reported that MDH can undergo a variant of inhibition[82] that results in the slowing-down of the oxidation/reduction of one of the two substrate/product redox couples. As a result of this observation, modeling of the filtering effect here is quite different from that for the earlier-encountered[31,53-58,65] "intensity filtering" systems. We develop an appropriate description and then verify it by applying the model to data for a system where the initial linear response, Scheme 1(b), is obtained by the biocatalytic action of another enzyme, glucose dehydrogenase, GDH.

**EXPERIMENTAL SECTION**

*Materials and instrumentation*

Malate dehydrogenase (MDH; E.C. 1.1.1.37) from porcine heart, glucose dehydrogenase (GDH; E.C. 1.1.1.47) from *Pseudomonas sp.*, β-nicotinamide adenine dinucleotide sodium salt ($NAD^+$), D-(+)-glucose, oxaloacetic acid, and 2-amino-2-hydroxymethyl-propane-1,3-diol (Tris-buffer) were obtained from Sigma-Aldrich and used without further purification. All experiments were carried out in ultrapure water (18.2 MΩ·cm; Barnstead NANOpure Diamond). A Shimadzu UV-2450 UV–Vis spectrophotometer with 1 mL poly(methyl methacrylate) (PMMA) cuvettes, was used for all optical kinetic measurements.

*Experimental system and procedures*

The biocatalytic processes probed in the experimental setup are schematically shown in Scheme 2. All experiments were performed at room temperature, 23 ± 2 °C. *Operation of the biocatalytic system without filtering:* Glucose (varied from 0 to 8 mM), GDH (326 mU/mL),

MDH (400 mU/mL), and NAD$^+$ (1.5 mM), were combined in Tris-buffer (0.5 M, pH 7.4) and used to generate NADH whose absorbance at 340 nm was monitored for 600 seconds. *Operation of biocatalytic system with filtering:* The above experiments were repeated in the presence of oxaloacetic acid (0.3 mM). Note that oxaloacetic acid is ionized to the oxaloacetate ion in the buffer solution, pH 7.4, and therefore it is referenced in the text as "oxaloacetate."

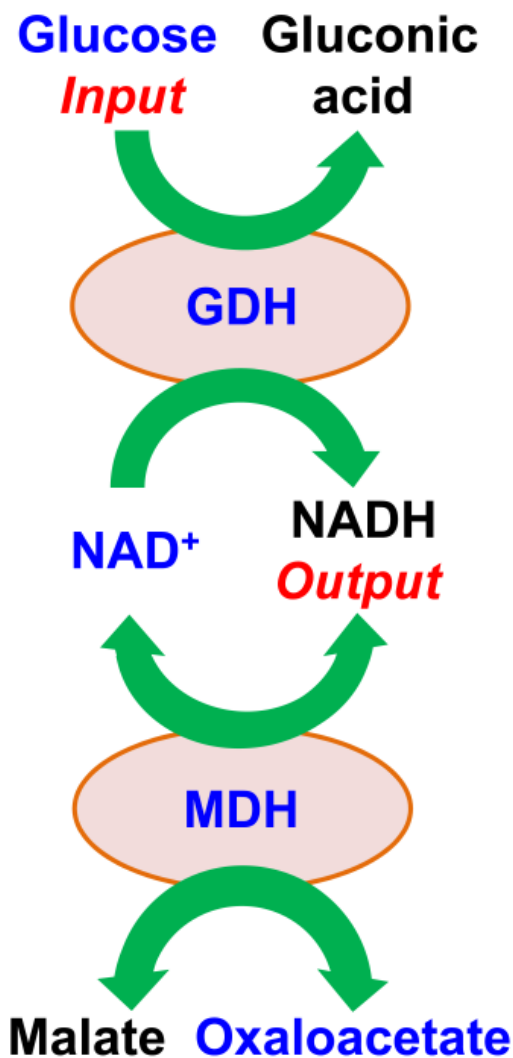

**Scheme 2.** The schematics of the enzymatic processes in the considered biocatalytic cascade. The chemicals which are initially in the system with filtering (see text) are marked in blue. The double-arrows schematically highlight the fact that the function of MDH is reversible.

**THEORETICAL SECTION**

*Linear signal transduction followed by fast reversible deactivation of the output*

Full kinetic description of enzymatic processes requires in most cases numerous parameters (rate constants) to describe the functioning of each enzyme. This aspect of the modeling will be further discussed later in this section. In this subsection, we consider a simple model with a minimal number of parameters that could be proposed to describe the effect on a linear response — of the type shown in Scheme 1(b) — of an added process that affects the output product, the concentration of which will be denoted $P(t)$, by rapidly converting it to and equilibrating it with another compound that is inert as far as contributing to the output signal measurement is concerned. Even though our ultimate conclusion will be that this simple description is not adequate for the present system, the model itself is important to study because adding fast, reversible processes of this sort by chemical or biochemical means can be done in numerous ways.

Here the first enzyme in the cascade, GDH, is utilized as a biocatalyst in the kinetic regime typical for many uses of all enzymes, i.e., both of its input chemicals (substrates), glucose and $NAD^+$, are provided with the initial concentrations large enough to have the products of the reaction generated with a constant rate as functions of time. For our product of interest, NADH, we thus assume that its concentration, $P(t)$, varies according to

$$\frac{dP}{dt} = RG, \qquad P(t_g) = RG t_g. \tag{1}$$

where in our case $R$ is a rate constant that can be fitted from the data, whereas $G$ is the initial concentration of glucose, which is our input at time $t = 0$, varied from 0 to 8 mM. We note that other reagents in the present system have fixed initial concentrations. The linear behavior in time applies for all but the smallest inputs, $G$, and it breaks down for very short times as well as for very long times on the time-scales of the experiments, which involved measuring the output in steps of fractions of seconds, for up to 600 s.

The second enzyme in the cascade, MDH, is also used in the regime of plentiful supply of its second substrate (other than our output), oxaloacetate, provided with a large initial concentration. Since its functioning is reversible (we consider details later), we could attempt to describe the kinetics of the present system by the effective processes

$$G \xrightarrow{R} P, \qquad P \underset{r_-}{\overset{r_+}{\rightleftarrows}} M. \tag{2}$$

We note that MDH actually oxidizes NADH to NAD$^+$, which is then our "inert" compound, but since NAD$^+$ is already present in the system in a large quantity, the variation of its concentration has little effect on the reverse process. However, malate, denoted, $M(t)$, see Scheme 2, which is not present initially, directly (and for simplicity we assume linearly) affects the reverse process rate. The appeal of the present model is not in its accuracy, but its simplicity and the fact that the resulting rate equations can be solved in closed form,

$$\frac{dP}{dt} = RG - r_+P + r_-M, \qquad \frac{dM}{dt} = r_+P - r_-M, \tag{3}$$

$$P(t) = RG\left\{\frac{r_+[1-e^{-(r_++r_-)t}]}{(r_++r_-)^2} + \frac{r_-t}{r_++r_-}\right\}. \tag{4}$$

One would assume that adding a fast reversible process that deactivates a part of the product, up to a fraction that corresponds to the concentrations of the rate-limiting chemicals for which that reversible process equilibrates, might have some "filtering" effect. However, the result obtained in Eq. (4) suggests that there is no filtering at all. Indeed, the dependence of the product $P(t_g)$ on the input, $G$, remains linear for any fixed "gate time" $t_g$, with a reduced slope (means, with loss of intensity). The original time-dependence, cf. Eq. (1), is linear in both $G$ and $t_g$. However, with the added process the time dependence and the input dependence are no longer thus related. As illustrated in Scheme 3, at small times the rate of the product output is unchanged because then the added process is not active. For large times a reduced rate, $RGr_-/(r_++r_-)$, is approached, as shown in the scheme.

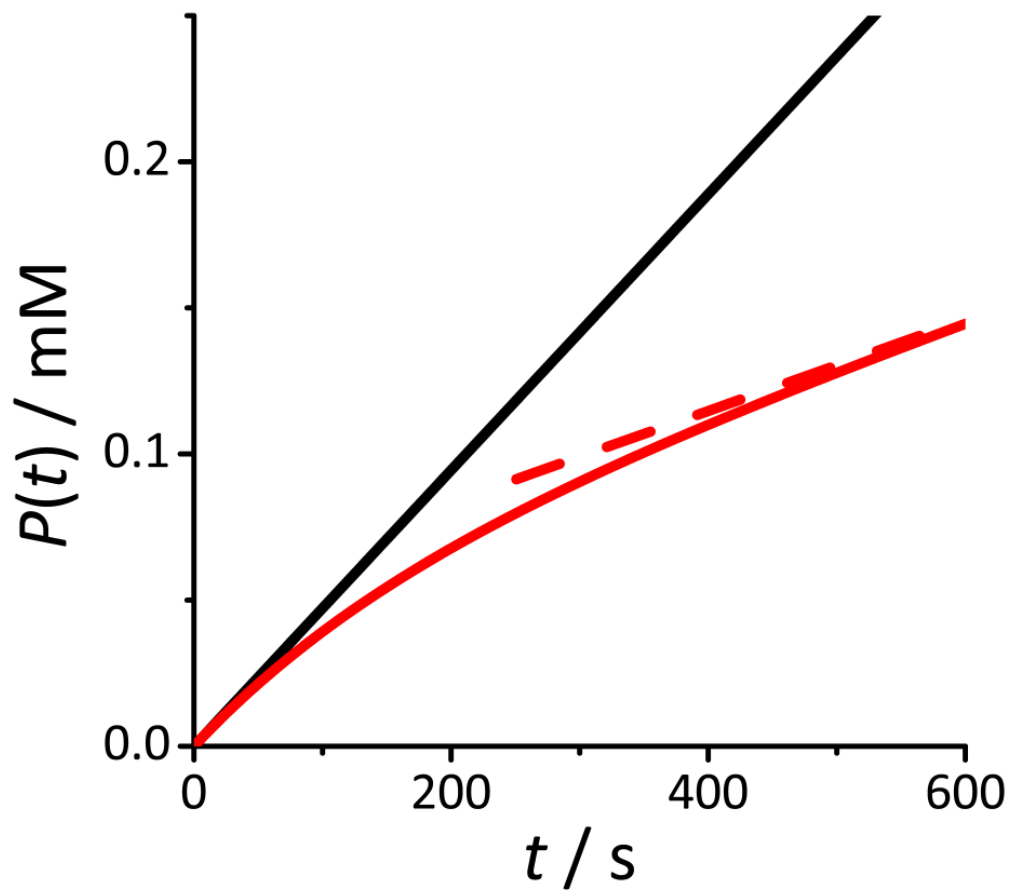

**Scheme 3.** Time dependence of the NADH concentration for arbitrarily selected typical parameter values ($G = 7$ mM, $R = 6.75 \times 10^{-6}$ s$^{-1}$, $r_+ = 2r_- = 4 \cdot 10^{-3}$ s$^{-1}$) with (the red curve) and without (the black straight line) the added fast reversible "output deactivation" process. The dashed line is the asymptotic slope (rate) for large times.

We thus reach an interesting conclusion that the experimentally observed[25] change from the linear to threshold response, Scheme 1(b) to 1(d), is due to more complicated kinetic mechanisms than the simple one just considered. The motivation for the present study has been to identify the origin of the observed effect, which turns out to be connected to an interesting kinetic property of the functioning of the enzyme MDH, as will be described in the rest of this section. We note that the present model offers a general conclusion that adding a fast, reversible process of deactivation of the input by equilibrating it with another species cannot in itself result in threshold type (at low inputs) intensity filtering. Examples[31,53-65] when such an approach worked have always involved the absence of equilibration by a limitation on how much of the other species can be produced (imposed by the process requiring some other, limited-supply chemical as an input).

*A model for the MDH kinetics with inhibition*

All enzymes have rather complicated kinetic mechanisms that involve the formation of complexes with their substrates, then processes involving these complexes, etc., in most cases resulting in the final restoration of the enzyme at the end of the cascade, when products are released. This includes our first enzyme, GDH, the mechanism of action of which is relatively standard,[84-86] but would require several rate constants to fully model. The second enzyme, MDH, has a rather complicated mechanism of action[82,87-89] as far as the number of intermediate complexes is concerned. The mechanism is in fact not fully studied. MDH can form complexes[82] with all four of the relevant substrates for its direct (NADH and oxaloacetate) or reverse (NAD$^+$ and malate) functioning, and then form triple-complexes in which the actual redox-pair conversions occur. Modeling[90] of such processes in their full kinetic detail would require at least 18 rate constants. This illustrates why it is important to use simplified (few-parameter) kinetic models for a semi-qualitative description of the response in sensors and biomolecular computing applications. Such approaches[21,53] usually involve setting up an effective phenomenological rate equation description that captures the main pathways of the involved processes.

The output product, NADH, denoted *P* for brevity, once it is generated by the GDH process, activates all the "direct" complex-formation and redox conversions involving MDH. As a result, not only is NADH partially converted back to NAD$^+$, to be denoted *N*, but also the concentration of malate, denoted *M*, builds up. This can gradually also activate the "reverse" processes involving MDH, driving the system towards equilibration. Measurements have been reported in the literature[82] that, as the relative concentration of malate as compared to oxaloacetate is increased, the redox inter-conversion rate NADH ↔ NAD$^+$ actually slows down, whereas the inter-conversion rate oxaloacetate (to be denoted *O*) ↔ malate increases. While at first this might look paradoxical, the likely explanation is as follows. Most of the enzyme, denoted *E*, gets "stuck" in the complexes *EP* and *EN* (as well as in more complicated complexes,

*ENM* and *EPO*). The fast inter-conversion oxaloacetate ↔ malate ($O \leftrightarrow M$) is accompanied (one might say, biocatalyzed) by the inter-conversion $EP \leftrightarrow EN$. This interesting mechanism can be entirely kinetic or can also be caused by malate inhibiting[82] some of the reaction pathways. This is not well-known, and such a study is outside the scope of the present work. It is important to emphasize that despite the earlier experimental evidence,[82] the considered mechanism of the redox pair $EP \leftrightarrow EN$ replacing $P \leftrightarrow N$ as the one accompanying the redox inter-conversion $O \leftrightarrow M$ as the concentration of malate builds up, is a conjecture. In fact, the observation that this assumption leads to modeling that fits the data, as reported in the next section, provides an additional support to this conjecture.

We model this effect phenomenologically, with a minimal possible number of parameters. Considering that oxaloacetate is supplied in large quantity, we can ignore its depletion. We assume that the concentration of malate that would correspond to steady state is $M_0$. We then write the rate equation of the linear supply of the product, cf. Eq. (1), but now also with the added term for the depletion of the product,

$$\frac{dP}{dt} = RG - K(M_0 - M)P = -KP^2 - K(M_0 - Rt)P + RG, \qquad (5)$$

where $K$ is the rate constant for the decrease in the amount of the product, $P$, due to the initially active mechanism, which, however, is gradually replaced by the mechanism involving $EP \leftrightarrow EN$ as $M$ increases from 0 to $M_0$. This assumes that the relative rates of the two mechanisms are directly proportional to $M_0 - M$ and $M$, respectively. The second expression in Eq. (5) was obtained by using $M(t) = RGt - P(t)$. The resulting equation for $P(t)$ is solved by

$$P(t) = RGt - M_0 + \frac{M_0 e^{-K\left(\frac{1}{2}RGt - M_0\right)t}}{1 + KM_0 \int_0^t e^{-K\left(\frac{1}{2}RG\tau - M_0\right)\tau} d\tau}, \qquad (6)$$

or, equivalently,

$$P(t) = RGt - M_0 + \frac{2\sqrt{KRG}M_0 e^{\frac{Kt}{2}(2M_0-RGt)}}{\sqrt{2\pi}KM_0 e^{\frac{KM_0^2}{2RG}}\left[\text{erf}\left(\sqrt{\frac{K}{2RG}}M_0\right)-\text{erf}\left(\sqrt{\frac{K}{2RG}}(M_0-RGt)\right)\right]+2\sqrt{KRG}}. \qquad (7)$$

This expression provides the dependence of $P(t_g)$ on $G$, of the type shown in Scheme 1(d), and its application for data fitting is described in the next section.

**RESULTS AND DISCUSSION**

*Considerations for data fitting*

Our primary objective in this work has been to demonstrate that a form of an inhibition mechanism in the functioning of MDH results in threshold filtering of the type that converts a linear response shown in Scheme 1(b) to that of Scheme 1(d). In order to confirm the validity of the proposed model, Eqs. (5-7), we will now use it to fit experimental data for the system sketched in Scheme 2. The time-dependent data were collected for times up to 600 s, in steps of a fraction of a second, for the following initial glucose concentrations, $G = 0.25, 0.5, 1, 2, 3, 4, 5, 6, 7$ and 8 mM. All the other chemicals that are initially present in the system were at the same initial concentrations for all the experiments without (no oxaloacetate) and with (fixed initial amount of oxaloacetate added) filtering.

Generally, the obtained data are rather noisy, fluctuating at least ±10% (of the maximum signal) seen as spread in the data and also as variations between experimental realizations, most of which were repeated 3 to 4 times for this estimate. The latter can be partially attributed to variations in and also degradation of the enzyme activity. This is typical for such enzymatic cascades, and underscores the value of developing few-parameter models and the preference[30-44] for a linear response, Scheme 1(b). Therefore, in our presentation of the results we focus on a single set of data, to minimize the effect of the enzyme degradation, taken continuously by varying (increasing) the input signal (glucose) first without the "filter" and then with it. We primarily report results for data collected for glucose values $G = 2, 3, 4, 5, 6, 7, 8$ mM, because the data for the lower glucose values are particularly noisy, and also the time-dependence of the output is not even approximately linear, which means that the simplest approximation for the non-filtered response, Eq. (1), is not accurate (better phenomenological modeling approaches are possible,[21,53] but would require more adjustable parameters).

The present model can provide imprecise data fitting for three main reasons: first, the data are noisy; second, the model itself is approximate, with effective rate parameters lumping together many actual chemical steps, as described in the preceding section; third, some of the data are taken in the regime for which the model might not be designed, specifically, when the first enzyme, GDH, is not supplied with enough glucose to "drive" its response to be linear, etc. With these reservations in mind, we carried out data fitting of the time dependence and also of the input (glucose) dependence at fixed times.

*Results of data fitting*

Time-dependence of two typical data sets is illustrated in Fig. 1. We note the deviations from the linear response for larger times and for smaller glucose inputs. The model curves shown will be described shortly. Figure 2 illustrates input-dependence at two different "gate times" $t_g$. The model curves here will also be described later. Let us first consider the (approximate) linear response without filtering. Figure 3 shows data sets for glucose inputs 2 mM and larger, for times up to $t = 200$ s. The data for the smaller $G$ values clearly show the inherent noise. Note that the differences between realizations with different $G$ values, seen as the dots fluctuating about the general trend in Fig. 2, and between repeats with the same $G$ values (not shown here) are much larger, because the time-variation (Fig. 3) is strongly correlated within each set.

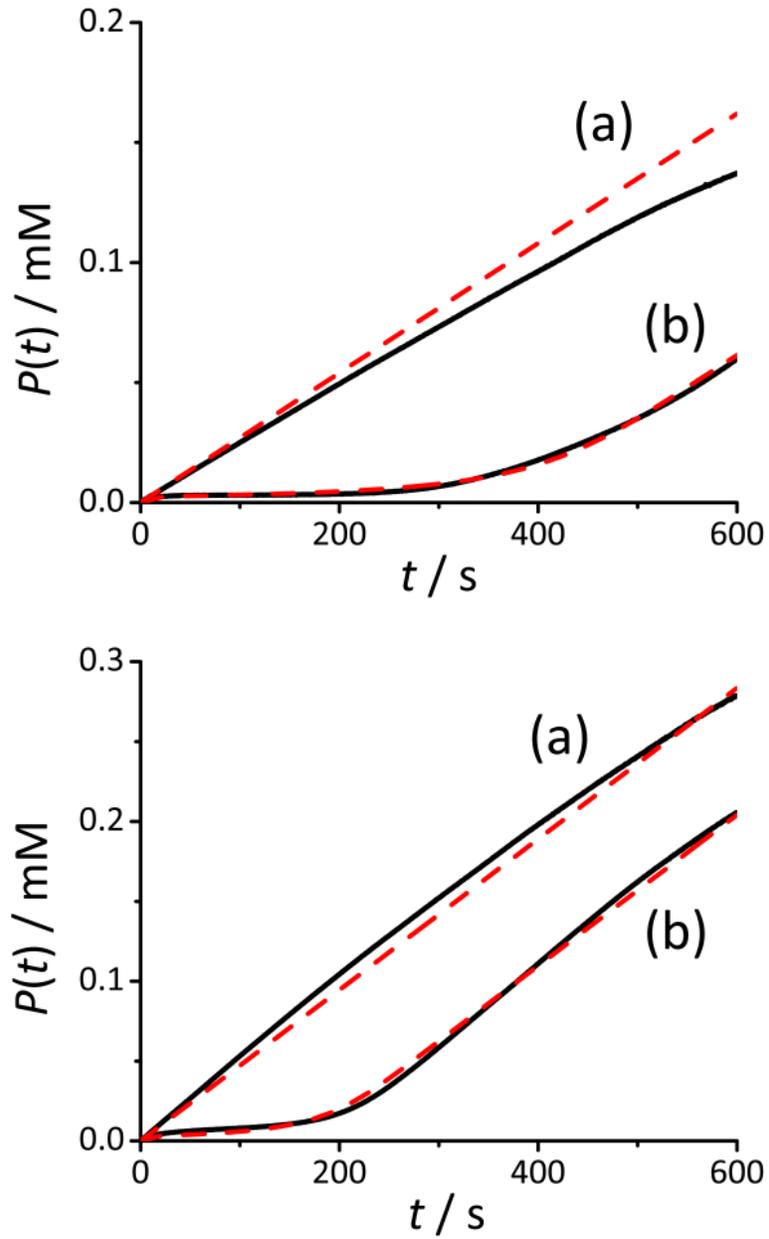

**Figure 1.** Top panel: Measured time dependence for input $G = 4$ mM (a set of experimental points that overlap to practically merge into noisy-looking solid lines). (a) without filtering; (b) with filtering. Bottom panel: The same for input $G = 7$ mM. The dashed lines show model fits as described in the text.

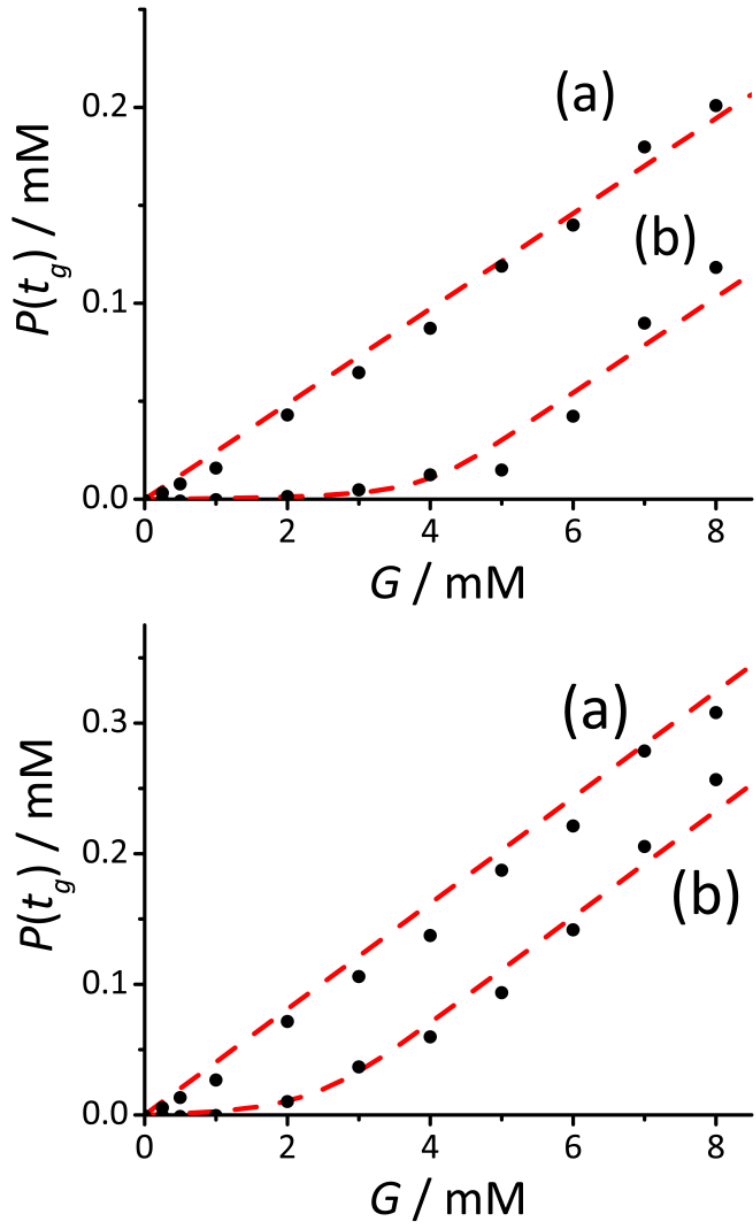

**Figure 2.** Top panel: Measured glucose dependence for fixed time $t = t_g = 360$ s (shown as dots). (a) without filtering; (b) with filtering. Bottom panel: The same for $t_g = 600$ s. The dashed lines show model fits as described in the text.

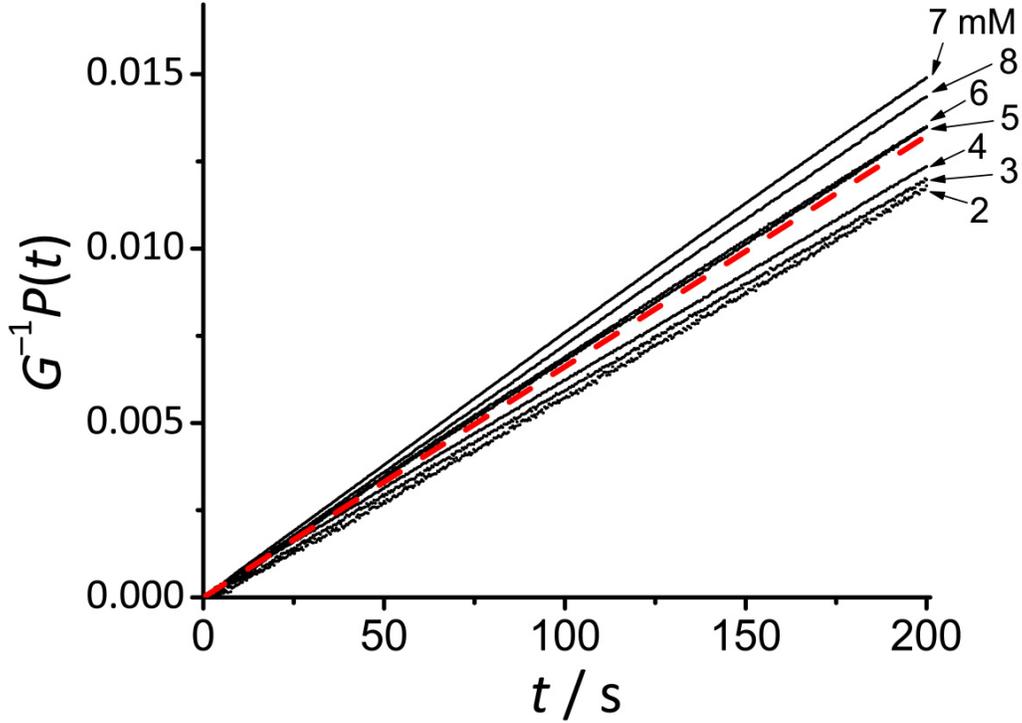

**Figure 3.** RMS fit of the experimental data for different initial glucose concentrations. Top to bottom: data sets of overlapping points correspond to $G$ = 7, 8, 6, 5, 4, 3, 2 mM, as labeled in the figure. The dashed straight line has slope $R = 6.75 \times 10^{-5}$ s$^{-1}$, which is the value obtained by fitting all the shown data together.

Note that the non-filtered data for the times and glucose inputs shown are rather linear. The quantity $P(t)/G$ plotted in Fig. 3, should actually be a single straight line if the model assumptions leading to Eq. (1) are correct. However, as mentioned earlier, noise and systematic variations are present. Still, the data in Fig. 3 as a whole can be well fitted by a straight line with the representative (for all data sets) slope $R = 6.75 \times 10^{-5}$ s$^{-1}$. The same line is also shown in Fig. 1 for a larger time span, as the model fit for the non-filtered case, illustrating the approximate linear behavior for the shown data sets, as well as various degrees of deviation for larger times. The trends for the other data sets without filtering were similar. In Fig. 2 the straight lines labelled (a), drawn as the model fits for the non-filtered case, correspond to the same rate constant, $R$, values. The slopes of these lines are $Rt_g$.

We note that data fitting without filtering does not actually probe the validity of the proposed kinetic mechanism for filtering. Rather, it checks the assumptions leading to the simplified linear response model, Eq. (1). Therefore, for this part of the data analysis we utilize a single overall-slope rate constant, $R$, value. We will use this value also for fitting data with filtering added. In the latter case we will explore how well the model equations, Eqs. (6-7), fit specific data sets. We then also propose representative parameter values for $M_0$ and $K$ that can be used for semi-quantitative fitting of the present data. With proper care and adjustment these parameter values, $R$, $M_0$, $K$ can by useful for fitting data for other systems[25] with the same MDH filtering mechanism and with an approximately linear output of the product, NADH, without filtering. In Fig. 1, the fitted curves, labelled (b), were obtained by using the earlier determined rate constant $R$, but the rate constant $K$ and the value of $M_0$ were both fitted from the data for that particular set of time dependence (for the specific value of $G$) by using a two-parameter RMS fit. The quality of the fits is quite good, with a clearly defined threshold behavior and linear increase in the output signal at later times, especially for the larger $G$ value shown. Similar behavior was found for other $G$ values, and the results for the parameters are summarized in Table 1. Similarly, data for several "gate times," from 300 to 600 s, were fitted, as illustrated by the curves labelled (b) in Fig. 2, and the same comments regarding the quality of the model description apply. The resulting values of the parameters are given in Table 2.

**Table 1.** Results of parameter fitting of time-dependent data sets for fixed $G$.

| $G$ (mM) | $M_0$ ($\mu M$) | $K$ ($mM^{-1}s^{-1}$) |
|---|---|---|
| 8 | 73 | 1.55 |
| 7 | 79 | 1.86 |
| 6 | 106 | 0.81 |
| 5 | 116 | 1.43 |
| 4 | 100 | 1.06 |
| 3 | 85 | 1.68 |
| 2 | 90 | 3.17 |

**Table 2.** Results of parameter fitting of input (glucose) dependent data sets for fixed $t_g$.

| $t_g$ (s) | $M_0$ (µM) | $K$ (mM$^{-1}$s$^{-1}$) |
|---|---|---|
| 300 | 87 | 5.28 |
| 360 | 92 | 2.48 |
| 420 | 93 | 1.61 |
| 480 | 95 | 0.93 |
| 540 | 93 | 0.62 |
| 600 | 92 | 0.43 |

The parameter values reported in Tables 1 and 2 reflect the fact that the data are rather noisy, as emphasized earlier, but also the property that the data sets are not large enough to determine both parameters accurately. The $K$ values in Table 2, and to a lesser extent the $M_0$ values in Table 1, illustrate this, and again underscore the earlier stressed fact that the use of as few parameters as possible in modeling data for such systems is important. Notwithstanding the noise in the data and in the fitted mean parameter values, we can define representative values and consider the quality of the data fits with these values used. Averaging the results for $M_0$ and $K$ in Table 1, which showed generally less spread than those in Table 2, we propose a set of representative parameters to supplement the earlier determined value of $R$, to yield

$$R = 6.75 \times 10^{-5} \text{ s}^{-1}, \quad K = 1.65 \text{ mM}^{-1}\text{s}^{-1}, \quad M_0 = 93 \text{ µM}. \tag{8}$$

We note that our primary application for the proposed model is to describe the glucose-dependence of the output for fixed "gate times." Figure 4 shows examples of the model curves drawn with the parameter set in Eq. (8), compared to the data, for two gate times somewhat different from those selected for Fig. 2. Unlike Fig. 2, however, the curves labelled (b) were not fitted using the specific data shown, but rather calculated with values from Eq. (8). Still, the quality of the data description remains satisfactory for the shown and other gate times from 300 up to 600 s.

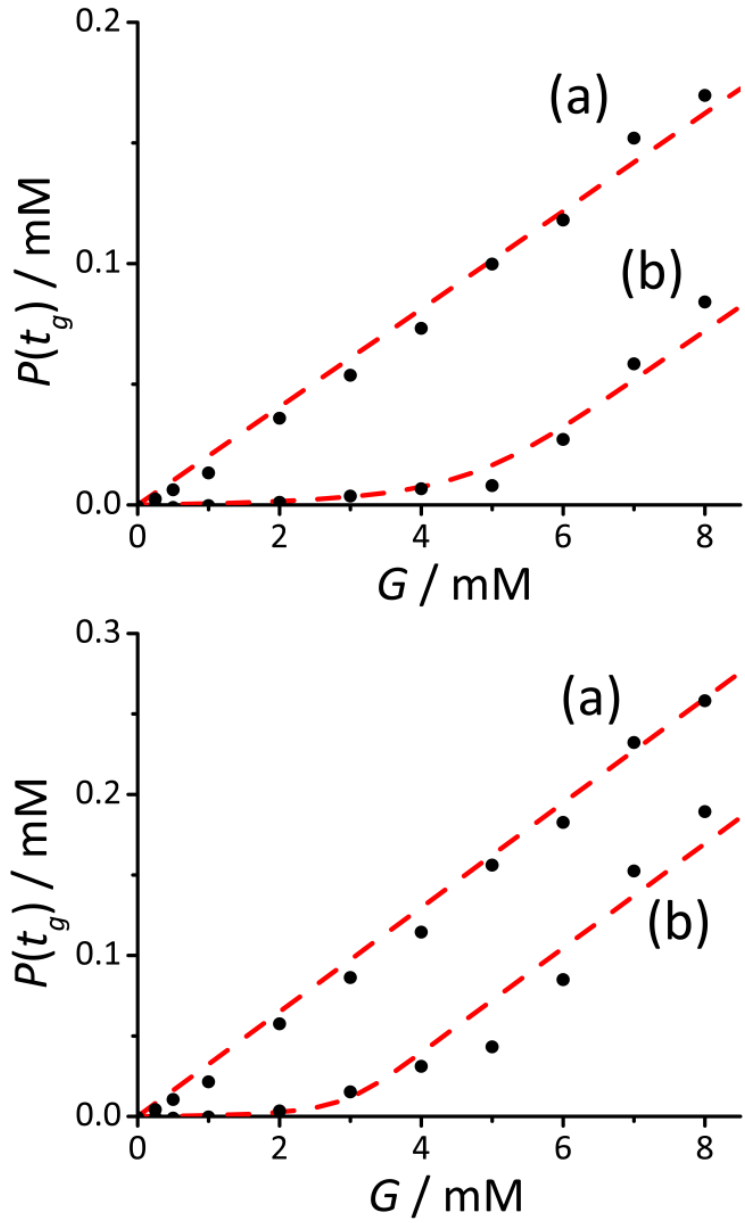

**Figure 4.** Top panel: Measured glucose-dependence for fixed time $t = t_g = 300$ s (shown as dots). (a) without filtering; (b) with filtering. Bottom panel: The same for $t_g = 480$ s. The dashed lines were drawn by using representative model parameters given in Eq. (8).

*Conclusion*

In summary, in this work we developed a theoretical understanding and simple few-parameter models explaining why fast reversible conversion of the product to another compound cannot on its own result in (bio)chemical "filtering." Filtering by the added processes catalyzed by the enzyme MDH, observed in earlier experiments, was then attributed to the enzyme's unusual mechanism of action. Experimental data for a system that linearly outputs the product that MDH can convert to another chemical was analyzed within the proposed model. Successful quantitative application of the model confirms the proposed mechanism for filtering. These findings will find applications in selecting other enzyme-catalyzed processes as potential "filtering" network steps in biomolecular cascades. Future work can study the "modularity" of such filtering by modeling the same filter process with other enzymatic systems[25,81] that can output the same product, NADH.

**ACKNOWLEDGEMENTS**

We thank Han Yan for collaboration on this project, and gratefully acknowledge funding of our research by the NSF, via award CBET-1066397. Y. Holade gratefully acknowledges the financial support from the French National Research Agency (ANR, ChemBio-Energy program) for his thesis project and from "le Conseil Regional Poitou-Charentes," "la Fondation Poitiers Université," and the doctoral school "Gay-Lussac" for his stay at Clarkson University.

# REFERENCES

(1) de Silva, A. P. Molecular Computing - A Layer of Logic. *Nature* **2008**, *454*, 417–418.

(2) de Silva, A. P. Molecular Logic and Computing. *Nature Nanotechnol.* **2007**, *2*, 399–410.

(3) Pischel, U.; Andreasson, J.; Gust, D.; Pais, V. F. Information Processing with Molecules-Quo Vadis? *ChemPhysChem* **2013**, *14*, 28–46.

(4) Pischel, U. Advanced Molecular Logic with Memory Function. *Angew. Chem. Int. Ed.* **2010**, *49*, 1356–1358.

(5) Benenson, Y. Biocomputers: From Test Tubes to Live Cells. *Molec. Biosys.* **2009**, *5*, 675–685.

(6) Benenson, Y. Biomolecular Computing Systems: Principles, Progress and Potential. *Nature Rev. Gen.* **2012**, *13*, 455–468.

(7) Stojanovic, M. N.; Stefanovic, D.; Rudchenko, S. Exercises in Molecular Computing. *Acc. Chem. Res.* **2014**, *47*, 1845–1852.

(8) Stojanovic, M. N.; Stefanovic, D. Chemistry at a Higher Level of Abstraction. *J. Comput. Theor. Nanosci.* **2011**, *8*, 434–440.

(9) Katz, E.; Privman, V. Enzyme-Based Logic Systems for Information Processing. *Chem. Soc. Rev.* **2010**, *39*, 1835–1857.

(10) Liu, Y.; Kim, E.; White, I. M.; Bentley, W. E.; Payne, G. F. Information Processing Through a Bio-Based Redox Capacitor: Signatures for Redox-Cycling. *Bioelectrochemistry* **2014**, *98*, 94–102.

(11) Jiang, H.; Riedel, M. D.; Parhi, K. K. Digital Signal Processing with Molecular Reactions. *IEEE Design Test Computers* **2012**, *29*, 21–31.

(12) Hillenbrand, P.; Fritz, G.; Gerland, U**.** Biological Signal Processing with a Genetic Toggle Switch. *PLOS ONE* **2013**, *8*, article # e68345.

(13) Bowsher, C. G. Information Processing by Biochemical Networks: A Dynamic Approach. *J. Royal Soc. Interface* **2011**, *8*, 186–200.

(14) Buisman, H. J.; ten Eikelder, H. M. M.; Hilbers, P. A. J.; Liekens, A. M. L. Computing Algebraic Functions with Biochemical Reaction Networks. *Artificial Life* **2009**, *15*, 5–19.

(15) Jia, Y. M.; Duan, R. X.; Hong, F.; Wang, B. Y.; Liu, N. N.; Xia, F. Electrochemical Biocomputing: A New Class of Molecular-Electronic Logic Devices. *Soft Matter* **2013**, *9*, 6571–6577.

(16) Zhou, M.; Dong, S. J. Bioelectrochemical Interface Engineering: Toward the Fabrication of Electrochemical Biosensors, Biofuel Cells, and Self-Powered Logic Biosensors. *Acc. Chem. Res.* **2011**, *44*, 1232–1243.

(17) Zhou, M.; Zhou, N.; Kuralay, F.; Windmiller, J. R.; Parkhomovsky, S.; Valdés-Ramírez, G.; Katz, E.; Wang, J. A Self-Powered "Sense-Act-Treat" System That Is Based on a Biofuel Cell and Controlled by Boolean Logic. *Angew. Chem. Int. Ed.* **2012**, *51*, 2686–2689.

(18) Katz, E. Bioelectronic Devices Controlled by Biocomputing Systems. *Isr. J. Chem.* **2011**, *51*, 132–140.

(19) Katz, E.; Wang, J.; Privman, M.; Halámek, J. Multi-Analyte Digital Enzyme Biosensors with Built-in Boolean Logic. *Anal. Chem.* **2012**, *84*, 5463–5469.

(20) Wang, J.; Katz, E. Digital Biosensors with Built-in Logic for Biomedical Applications – Biosensors Based on Biocomputing Concept. *Anal. Bioanal. Chem.* **2010**, *398*, 1591–1603.

(21) Privman, V. Control of Noise in Chemical and Biochemical Information Processing. *Isr. J. Chem.* **2011**, *51*, 118–131.

(22) Privman, V. Approaches to Control of Noise in Chemical and Biochemical Information and Signal Processing. Ch. 12 in: *Molecular and Supramolecular Information Processing – From Molecular Switches to Unconventional Computing.* Katz, E. (Ed.), Willey-VCH, Weinheim, **2012**, pages 281-303.

(23) Katz, E.; Privman, V.; Wang, J. Towards Biosensing Strategies Based on Biochemical Logic Systems, E. in: Proceedings of The Fourth International Conference on Quantum, Nano and Micro Technologies (ICQNM 2010), Ovchinnikov, V.; Privman, V. (Eds.), IEEE Computer Society Conference Publishing Services, Los Alamitos, California, **2010**, pages 1-9.

(24) Guz, N.; Halámek, J.; Rusling, J. F.; Katz, E. A Biocatalytic Cascade with Several Output Signals − Towards Biosensors with Different Levels of Confidence. *Anal. Bioanal. Chem.* **2014**, *406*, 3365-3370.

(25) Mailloux, S.; Guz, N.; Zakharchenko, A.; Minko, S.; Katz, E. Majority and Minority Gates Realized in Enzyme-Biocatalyzed Systems Integrated with Logic Networks and Interfaced with Bioelectronic Systems. *J. Phys. Chem. B* **2014**, *118*, 6775–6784.

(26) MacVittie, K.; Halámek, J.; Privman, V.; Katz, E. A Bioinspired Associative Memory System Based on Enzymatic Cascades. *Chem. Commun.* **2013**, *49*, 6962–6964.

(27) Halámek, J.; Bocharova, V.; Chinnapareddy, S.; Windmiller, J. R. Strack, G.; Chuang, M.-C.; Zhou, J.; Santhosh, P.; Ramirez, G. V.; Arugula, M. A. et. al. Multi-Enzyme Logic Network Architectures for Assessing Injuries: Digital Processing of Biomarkers. *Molec. Biosys.* **2010**, *6*, 2554–2560.

(28) Privman, V.; Arugula, M. A.; Halámek, J.; Pita, M.; Katz, E. Network Analysis of Biochemical Logic for Noise Reduction and Stability: A System of Three Coupled Enzymatic AND Gates. *J. Phys. Chem. B* **2009**, *113,* 5301–5310.

(29) Strack, G.; Ornatska, M.; Pita, M.; Katz, E. Biocomputing Security System: Concatenated Enzyme-Based Logic Gates Operating as a Biomolecular Keypad Lock. *J. Am. Chem. Soc.* **2008**, *130*, 4234–4235.

(30) Privman, V.; Zavalov, O.; Simonian, A. Extended Linear Response for Bioanalytical Applications Using Multiple Enzymes. *Anal. Chem.* **2013**, *85*, 2027–2031.

(31) Zavalov, O.; Domanskyi, S.; Privman, V. Simonian, A. Design of Biosensors with Extended Linear Response and Binary-Type Sigmoid Output Using Multiple Enzymes, in: *Proc. Conf. ICQNM 2013*, ThinkMind Online Publishing, Reed Hook, NY, **2013**, pp. 54–59.

(32) Rinken, T.; Rinken, P.; Kivirand, K. Signal Analysis and Calibration of Biosensors for Biogenic Amines in the Mixtures of Several Substrates. In: *Biosensors - Emerging Materials and Applications*, Serra, P. A. (Ed.); InTech.; Rijeka, Croatia, **2011**, pp. 3–16.

(33) Bakker, E. Electrochemical Sensors. *Anal. Chem.* **2004**, *76*, 3285–3298.

(34) Chen, X.; Zhu, J.; Tian, R.; Yao, C. Bienzymatic Glucose Biosensor Based on Three Dimensional Macroporous Ionic Liquid Doped Sol–Gel Organic–Inorganic Composite. *Sens. Actuat. B* **2012**, *163*, 272–280.

(35) Coche-Guérente, L.; Cosnier, S.; Labbé, P. Sol−Gel Derived Composite Materials for the Construction of Oxidase/Peroxidase Mediatorless Biosensors. *Chem. Mater*. **1997**, *9,* 1348–1352.

(36) De Benedetto, G. E.; Palmisano, F.; Zambonin, P. G. One-Step Fabrication of a Bienzyme Glucose Sensor Based on Glucose Oxidase and Peroxidase Immobilized onto a Poly(pyrrole) Modified Glassy Carbon Electrode. *Biosens. Bioelectron*. **1996**, *11*, 1001–1008.

(37) Delvaux, M.; Walcarius, A.; Demoustier-Champagne, S. Bienzyme HRP–GOx-Modified Gold Nanoelectrodes for the Sensitive Amperometric Detection of Glucose at Low Overpotentials. *Biosens. Bioelectron*. **2005**, *20*. 1587–1594.

(38) Ferri, T.; Maida, S.; Poscia, A.; Santucci, R. A Glucose Biosensor Based on Electro-Enzyme Catalyzed Oxidation of Glucose Using a HRP-GOD Layered Assembly. *Electroanalysis* **2001**, *13*, 1198–1202.

(39) Nikitina, O.; Shleev, S.; Gayda, G.; Demkiv, O.; Gonchar, M.; Gorton, L.; Csoregi, E.; Nistor, M. Bi-Enzyme Biosensor Based on $NAD^+$- and Glutathione-Dependent Recombinant Formaldehyde Dehydrogenase and Diaphorase for Formaldehyde Assay. *Sens. Actuat. B* **2007**, *125*, 1–9.

(40) Tian, F. M.; Zhu, G. Y. Bienzymatic Amperometric Biosensor for Glucose Based on Polypyrrole/Ceramic Carbon as Electrode Material. *Anal. Chim. Acta* **2002**, *451*, 251–258.

(41) Zhu, L.; Yang, R.; Zhai, J.; Tian, C. Bienzymatic Glucose Biosensor Based on Co-Immobilization of Peroxidase and Glucose Oxidase on a Carbon Nanotubes Electrode. *Biosens. Bioelectron.* **2007**, *23*, 528–535.

(42) Simonian, A. L.; Badalian I. E.; Smirnova, I. P.; Berezov, T. T. Metabolite Alternative Splitting by Different Enzymes. In: *Biochemical Engineering-Stuttgart*; Reuss, M. (Ed.); G. Fisher Pub., Stuttgart-New York, **1991**, pp. 344–347.

(43) Simonian, A. L.; Khachatrian, G. E.; Tatikian, S. S.; Avakian, T. M.; Badalian, I. E. A Flow-Through Enzyme Analyzer for Determination of L-Lysine Concentration. *Biosens. Bioelectron*. **1991**, *6*, 93–99.

(44) Simonian, A. L.; Badalian, I. E.; Berezov, T. T; Smirnova, I. P; Khaduev, S. H. Flow-Injection Amperometric Biosensor Based on Immobilized L-Lysine-α-Oxidase for L-Lysine Determination. *Anal. Lett*. **1994**, *27*, 2849–2860.

(45) *Biomolecular Information Processing - From Logic Systems to Smart Sensors and Actuators*, Katz, E. (Ed.), Wiley-VCH, Weinheim, Germany, **2012**.

(46) *Unconventional Computation. Lecture Notes in Computer Science*, Calude, C. S.; Costa, J. F.; Dershowitz, N.; Freire, E.; Rozenberg, G. (Eds.), Vol. 5715, Springer, Berlin, **2009**.

(47) *Unconventional Computing*, Adamatzky, A.; De Lacy Costello, B.; Bull, L.; Stepney, S.; Teuscher, C. (Eds.), Luniver Press, Frome, UK, **2007**.

(48) *Molecular and Supramolecular Information Processing – From Molecular Switches to Unconventional Computing.* Katz, E. (Ed.), Willey-VCH, Weinheim, **2012**.

(49) Lederman, H.; Macdonald, J.; Stefanovic, D.; Stojanovic, M. N. Deoxyribozyme-Based Three-Input Logic Gates and Construction of a Molecular Full Adder. *Biochemistry* **2006**, *45*, 1194–1199.

(50) Baron, R.; Lioubashevski, O.; Katz, E.; Niazov, T.; Willner, I. Elementary Arithmetic Operations by Enzymes: A Model for Metabolic Pathway Based Computing. *Angew. Chem. Int. Ed.* **2006**, *45*, 1572–1576.

(51) MacVittie, K.; Halámek, J.; Katz, E. Enzyme-Based D-Flip-Flop Memory System. *Chem. Commun.* **2012**, *48*, 11742–11744.

(52) Pita, M.; Strack, G.; MacVittie, K.; Zhou, J.; Katz, E. Set-Reset Flip-Flop Memory Based on Enzyme Reactions: Towards Memory Systems Controlled by Biochemical Pathways. *J. Phys. Chem. B* **2009,** *113,* 16071–16076.

(53) Privman, V.; Zavalov, O.; Halámková, L.; Moseley, F.; Halámek, J.; Katz, E. Networked Enzymatic Logic Gates with Filtering: New Theoretical Modeling Expressions and Their Experimental Application. *J. Phys. Chem. B* **2013**, *117*, 14928–14939.

(54) Halámek, J.; Zavalov, O.; Halámková, L.; Korkmaz, S.; Privman, V.; Katz, E. Enzyme-Based Logic Analysis of Biomarkers at Physiological Concentrations: AND Gate with Double-Sigmoid "Filter" Response. *J. Phys. Chem. B* **2012**, *116*, 4457–4464.

(55) Zavalov, O.; Bocharova, V.; Privman, V.; Katz, E. Enzyme-Based Logic: OR Gate with Double-Sigmoid Filter Response. *J. Phys. Chem. B* **2012**, *116*, 9683–9689.

(56) Zavalov, O.; Bocharova, V.; Halámek, J.; Halámková, L.; Korkmaz, S.; Arugula, M. A.; Chinnapareddy, S.; Katz, E.; Privman, V. Two-Input Enzymatic Logic Gates Made Sigmoid by Modifications of the Biocatalytic Reaction Cascades. *Int. J. Unconv. Comput.* **2012**, *8*, 347–365.

(57) Bakshi, S.; Zavalov, O.; Halámek, J.; Privman, V.; Katz, E. Modularity of Biochemical Filtering for Inducing Sigmoid Response in Both Inputs in an Enzymatic AND Gate. *J. Phys. Chem. B* **2013**, *117*, 9857−9865.

(58) Privman, V.; Fratto, B. E.; Zavalov, O.; Halámek, J.; Katz, E. Enzymatic AND Logic Gate with Sigmoid Response Induced by Photochemically Controlled Oxidation of the Output. *J. Phys. Chem. B* **2013**, *117*, 7559–7568.

(59) Rafael, S. P.; Vallée-Bélisle, A.; Fabregas, E.; Plaxco, K.; Palleschi, G.; Ricci, F. Employing the Metabolic "Branch Point Effect" to Generate an All-or-None, Digital-Like Response in Enzymatic Outputs and Enzyme-Based Sensors. *Anal. Chem.* **2012**, *84*, 1076–1082.

(60) Vallée-Bélisle, A.; Ricci, F.; Plaxco, K. W. Engineering Biosensors with Extended, Narrowed, or Arbitrarily Edited Dynamic Range. *J. Am. Chem. Soc.* **2012**, *134*, 2876−2879.

(61) Kang, D.; Vallée-Bélisle, A.; Plaxco, K. W.; Ricci, F. Re-engineering Electrochemical Biosensors to Narrow or Extend Their Useful Dynamic Range. *Angew. Chem. Int. Ed.* **2012**, *51*, 6717–6721.

(62) Halámek, J.; Zhou, J.; Halámková, L.; Bocharova, V.; Privman, V.; Wang, J.; Katz, E. Biomolecular Filters for Improved Separation of Output Signals in Enzyme Logic Systems Applied to Biomedical Analysis, *Anal. Chem.* **2011**, *83*, 8383–8386.

(63) Pita, M.; Privman, V.; Arugula, M. A.; Melnikov, D.; Bocharova, V.; Katz, E. Towards Biochemical Filter with Sigmoidal Response to pH Changes: Buffered Biocatalytic Signal Transduction. *Phys. Chem. Chem. Phys.* **2011**, *13*, 4507–4513.

(64) Privman, V.; Halámek, J.; Arugula, M. A.; Melnikov, D.; Bocharova, V.; Katz, E. Biochemical Filter with Sigmoidal Response: Increasing the Complexity of Biomolecular Logic. *J. Phys. Chem. B* **2010**, *114*, 14103–14109.

(65) Domanskyi, S.; Privman, V. Design of Digital Response in Enzyme-Based Bioanalytical Systems for Information Processing Applications. *J. Phys. Chem. B* **2012**, *116*, 13690–13695.

(66) Unger, R.; Moult, J. Towards Computing with Proteins *Proteins* **2006**, *63*, 53–64.

(67) Margolin, A. A.; Stojanovic, M. N. Boolean Calculations Made Easy (for Ribozymes). *Nature Biotechnol.* **2005**, *23*, 1374–1376.

(68) Win, M. N.; Smolke, C. D. Higher-Order Cellular Information Processing with Synthetic RNA Devices. *Science* **2008**, *322*, 456–460.

(69) Rinaudo, K.; Bleris, L.; Maddamsetti, R.; Subramanian, S.; Weiss, R.; Benenson, Y. A Universal RNAi-Based Logic Evaluator That Operates in Mammalian Cells. *Nature Biotechnol.* **2007**, *25*, 795–801.

(70) Simpson, M. L.; Sayler, G. S.; Fleming, J. T.; Applegate, B. Whole-Cell Biocomputing. *Trends Biotechnol.* **2001**, *19*, 317–323.

(71) Li, Z.; Rosenbaum, M. A.; Venkataraman, A.; Tam, T. K.; Katz, E.; Angenent, L. T. Bacteria-Based AND Logic Gate: A Decision-Making and Self-Powered Biosensor. *Chem. Commun.* **2011**, *47*, 3060–3062.

(72) Chuang, M.-C.; Windmiller, J. R.; Santhosh, P.; Valdés-Ramírez, G.; Katz, E.; Wang, J. High-Fidelity Simultaneous Determination of Explosives and Nerve Agent Threats via a Boolean Biocatalytic Cascade. *Chem. Commun.* **2011**, *47*, 3087–3089.

(73) Halámek, J.; Windmiller, J. R.; Zhou, J.; Chuang, M.-C.; Santhosh, P.; Strack, G.; Arugula, M. A.; Chinnapareddy, S.; Bocharova, V.; Wang, J.; et. al. Multiplexing of Injury Codes for the Parallel Operation of Enzyme Logic Gates. *Analyst* **2010**, *135*, 2249–2259.

(74) Xia, F.; Zuo, X. L.; Yang, R. Q.; White, R. J.; Xiao, Y.; Kang, D.; Gong, X. O.; Lubin, A. A.; Vallee-Belisle, A.; Yuen, J. D.; et. al. Label-Free, Dual-Analyte Electrochemical Biosensors: A New Class of Molecular-Electronic Logic Gates. *J. Am. Chem. Soc.* **2010**, *132*, 8557–8559.

(75) Mailloux, S.; Halámek, J.; Katz, E. A Model System for Targeted Drug Release Triggered by Biomolecular Signals Logically Processed Through Enzyme Logic Networks. *Analyst* **2014**, *139*, 982–986.

(76) Katz, E.; Bocharova, V.; Privman, M. Electronic Interfaces Switchable by Logically Processed Multiple Biochemical and Physiological Signals. *J. Mater. Chem.* **2012**, *22*, 8171–8178.

(77) Privman, M.; Tam, T. K.; Bocharova,V.; Halámek, J.; Wang, J.; Katz, E. Responsive Interface Switchable by Logically Processed Physiological Signals – Towards "Smart" Actuators for Signal Amplification and Drug Delivery. *ACS Appl. Mater. Interfaces* **2011**, *3*, 1620–1623.

(78) *Electrochemical Sensors, Biosensors and their Biomedical Applications*. Zhang, X.; Ju, H.; Wang J. (Eds.), Academic Press, NY, **2008**.

(79) Poghossian, A.; Malzahn, K.; Abouzar, M. H.; Mehndiratta, P.; Katz, E.; Schöning, M. J. Integration of Biomolecular Logic Gates with Field-Effect Transducers. *Electrochim. Acta* **2011**, *56*, 9661–9665.

(80) Krämer, M.; Pita, M.; Zhou, J.; Ornatska, M.; Poghossian, A.; Schöning, M. J.; Katz, E. Coupling of Biocomputing Systems with Electronic Chips: Electronic Interface for Transduction of Biochemical Information. *J. Phys. Chem. C* **2009**, *113*, 2573–2579.

(81) Mailloux, S.; Zavalov, Nataliia Guz, N.; Katz, E.; Bocharova, V. Enzymatic Filter for Improved Separation of Output Signals in Enzyme Logic Systems Towards 'Sense and Treat' Medicine. *Biomater. Sci.* **2014**, *2*, 184–191.

(82) Silverstein, E.; Sulebele, G. Catalytic Mechanism of Pig Heart Mitochondrial Malate Dehydrogenase Studied by Kinetics at Equilibrium. *Biochemistry* **1969**, *8*, 2543–2550.

(83) Marangoni, A. G. *Enzyme Kinetics. A Modern Approach*. John Wiley & Sons, Inc., New York, **2003**.

(84) Ohshima, T.; Ito, Y.; Sakuraba, H.; Goda, S.; Kawarabayasi, Y. The *Sulfolobus Tokodaii* Gene ST1704 Codes Highly Thermostable Glucose Dehydrogenase. *J. Molec. Catal. B* **2003**, *23*, 281–289.

(85) Strecker, H. J.; Korkes, S. Glucose Dehydrogenase. *J. Biol. Chem.* **1952**, *196*, 769–784.

(86) Bhaumik, S. R.; Sonawat, H. M. Kinetic Mechanism of Glucose Dehydrogenase from *Halobacterium Salinarum*. *Ind. J. Biochem. Biophys.* **1999**, *36*, 143–149.

(87) Cunningham, M. A.; Ho, L. L.; Nguyen, D. T.; Gillilan, R. E.; Bash, P. A. Simulation of the Enzyme Reaction Mechanism of Malate Dehydrogenase. *Biochemistry* **1997**, *36*, 4800–4816.

(88) Wolfe, R. G.; Neilands, J. B. Some Molecular and Kinetic Properties of Heart Malic Dehydrogenase. *J. Biol. Chem.* **1956**, *221*, 61–70.

(89) Minárik, P.; Tomášková, N.; Kollárová, M.; Antalík, M. Malate Dehydrogenases – Structure and Function. *Gen. Physiol. Biophys.* **2002**, *21*, 265–257.

(90) Cha, S. A Simple Method for Derivation of Rate Equations for Enzyme-Catalyzed Reactions under the Rapid Equilibrium Assumption or Combined Assumptions of Equilibrium and Steady State. *J. Biol. Chem.* **1968**, *243*, 820–825.

**GRAPHICAL ABSTRACT (Table of Contents Image)**

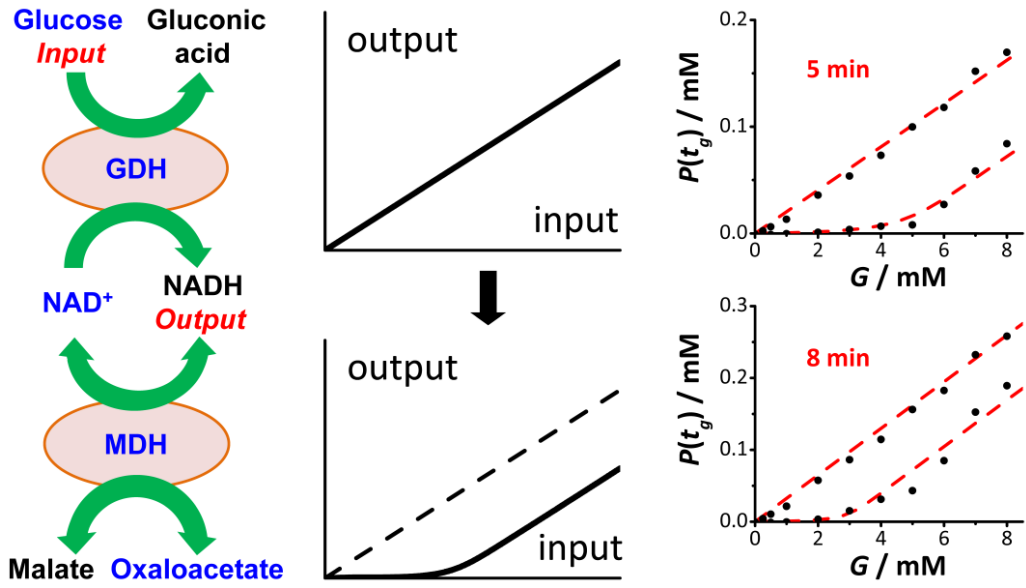

# Legend for the Journal Issue Cover

Threshold filtering by processes catalyzed by malate dehydrogenase is explained in terms of a model that incorporates an inhibition of this enzyme that has a reversible mechanism of action. Experimental data for a system in which the output signal is produced by biocatalytic processes of the enzyme glucose dehydrogenase are analyzed to verify the model's validity.

**Model of Threshold Filtering in an Enzymatic Cascade for Bioanalytical and Biocomputing Applications**
(see page 12435)

$$P(t) = RGt - M_0 + \frac{M_0 e^{-K\left(\frac{1}{2}RGt - M_0\right)t}}{1 + KM_0 \int_0^t e^{-K\left(\frac{1}{2}RG\tau - M_0\right)\tau} d\tau}$$

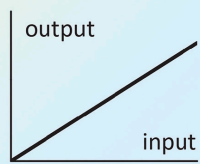

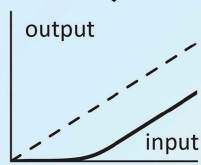

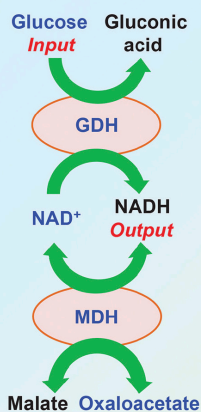

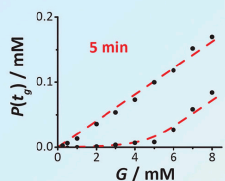

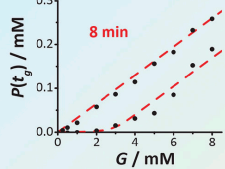

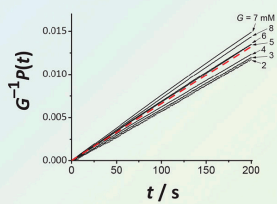

$$P(t) = RGt - M_0 + \frac{2\sqrt{KRG}\, M_0 e^{\frac{Kt}{2}(2M_0 - RGt)}}{\sqrt{2\pi} KM_0 e^{\frac{KM_0^2}{2RG}}\left[\mathrm{erf}\left(\sqrt{\frac{K}{2RG}}M_0\right) - \mathrm{erf}\left(\sqrt{\frac{K}{2RG}}(M_0 - RGt)\right)\right] + 2\sqrt{KRG}}$$

**BIOPHYSICAL CHEMISTRY, BIOMATERIALS, LIQUIDS, AND SOFT MATTER**